\documentstyle[12pt,a41,epsfig]{article}
\newcommand{\beq}{\begin{equation}}
\newcommand{\eeq}{\end{equation}}
\newcommand{\bea}{\begin{eqnarray}}
\newcommand{\eea}{\end{eqnarray}}
\newcommand{\ab}{\overline{\alpha}_s}

\newcommand{\gsim}{\raisebox{-0.07cm}{$\, \stackrel{>}{{\scriptstyle
\sim}}\, $}}

\newcommand\GeV{\,\mbox{GeV}}

\newcommand\GAM{\mbox{\boldmath $\gamma$}}

\begin{document}
\setlength{\baselineskip}{0.52cm}
\begin{titlepage}

\begin{flushleft}
DESY 97-143 \hfill {\tt hep-ph/9707488} \\
WUE-ITP-97-22 \hfill July 1997 \\
\hfill (Revised 10/97)
\end{flushleft}

\vspace*{2.5cm}
\begin{center}
{\Large\bf On the Resummed Gluon Anomalous Dimension}

\vspace{3mm}
{\Large\bf and Structure Functions at Small \mbox{\boldmath $x$}}

\vspace{3cm}
{\large Johannes Bl\"umlein$^a$ and Andreas Vogt$^b$}

\vspace{2cm}
{\it $^a$DESY--Zeuthen, Platanenallee 6, D--15735 Zeuthen, Germany}\\
\vspace{0.3cm}
{\it $^b$Institut f\"ur Theoretische Physik, Universit\"at W\"urzburg}\\
{\it Am Hubland, D--97074 W\"urzburg, Germany}

\vspace{4cm}
{\bf Abstract}
\vspace{-0.2cm}
\end{center}
The impact of the recently evaluated `irreducible' contributions to 
the resummed next-to-leading logarithmic small-$x$ anomalous dimension 
$\gamma_{gg}$ is evaluated for the unpolarized parton densities and 
structure functions of the nucleon. These new terms diminish the gluon 
distribution and are found to overcompensate the enhancement caused 
by the resummed leading logarithmic small-$x$ anomalous dimension and 
the quarkonic contributions beyond next-to-leading order.

\vspace{0.7cm}
\begin{center}
\vspace{1cm}
Phys.\ Rev.\ {\bf D57} (1998) 1
\end{center}

\end{titlepage}

\noindent
The structure functions of the proton in unpolarized deep--inelastic
scattering (DIS) show a steep rise towards very small Bjorken-$x$
values, which becomes stronger with increasing resolution $Q^2$. This
behavior is a consequence both of the shape of the non-perturbative
quark singlet and gluon initial distributions -- $x\Sigma$ and $xg$
rise roughly like $x^{-0.2}$ for a starting scale $Q^2_0 \simeq 4
\GeV^2$ -- and of the form of the evolution kernels governing the
renormalization group equations of the mass singularities.
The anomalous dimensions $\gamma_{ij}$ for the evolution of the parton
densities, as well as the Wilson coefficients $C_n$ of the structure
functions, contain large logarithmic \mbox{small-$x$} contributions.
In order to arrive at a reliable theoretical framework at very small
$x$, the resummation of these terms may be necessary to all orders in 
the strong coupling $\alpha_s(Q^2)$. For the unpolarized singlet case 
considered here the dominant contributions take, in Mellin-$N$ space, 
the form $\alpha_s^{l+k}/(N-1)^k , \: l \geq 0$.

The resummation of $\gamma_{gg}$ and $\gamma_{gq}$ in the leading
small-$x$ approximation (L$x)$, $l=0$, was performed long ago
\cite{LIP1}. $\gamma_{gg}^{(0)}(N)$ is obtained as the solution of
\beq
\label{eq1}
 1 = \frac{\ab}{N - 1} \, \chi_0(\gamma_{gg}^{(0)})
\eeq
with $\ab = C_A \alpha_s(Q^2)/\pi$, $C_A = N_c = 3$,
$C_F = 4/3$ and
\beq
\label{eq2}
 \chi_0(\gamma) = 2 \psi(1) - \psi(\gamma) - \psi(1- \gamma) \: ,
\eeq
furthermore $\gamma_{gq}^{(0)} = (C_F/C_A)\, \gamma_{gg}^{(0)}$. The
quark anomalous dimensions $\gamma_{qq}$ and $\gamma_{qg}$, on the
other hand, receive contributions for $l\geq 1$ only. The $l=1$ terms 
were derived in ref.~\cite{CH}, together with the corresponding 
resummations for the coefficient functions $C_2$ and $C_L$. The large 
effects of these quantities on the small-$x$ behavior of the DIS 
structure functions were subsequently studied in detail [3--7]. In 
those investigations the small-$x$ resummation of the gluon anomalous 
dimension $\gamma_{gg}$ to 
next-to-leading order small-$x$ (NL$x$)
accuracy could not be taken into 
account. This resummation has now been performed for the quarkonic 
contributions~[8--11] proportional to the number of quark flavors 
$N_f$. Recently also the `irreducible' gluonic terms $\propto C_A$ 
have been derived~\cite{CC2}, i.e., those contributions which are 
energy--scale independent in the framework of ref.~\cite{LIP2} 
underlying that calculation. The corresponding terms of $\gamma_{gq}$, 
however, still remain to be determined.

In this 
note 
we investigate the impact of these new resummed 
contributions to $\gamma_{gg}$ on the evolution of the parton densities 
and the proton structure functions $F_2(x,Q^2)$ and $F_L(x,Q^2)$, for
the first time including calculated subleading terms into the 
renormalization group analysis. Hence the comparison of the results 
to the findings of previous studies~[3--7] should allow for improved 
estimates of the convergence of the small-$x$ resummation 
approximation, despite a fully quantitative, scheme--independent NL$x$ 
analysis not being possible at present.

As will be demonstrated below, the effect of the new contributions to 
the resummed anomalous dimension $\gamma_{gg}$ is very large and 
opposite to that of the previously known resummed terms. This implies, 
already at the present stage, considerable changes particularly for 
gluon--dominated quantities, which partly modify conclusions obtained 
in previous numerical investigations~\mbox{[3--6]}. A detailed account 
of the solution of the evolution equations in the presence of all-order 
anomalous dimensions and coefficient functions will be given in a 
forthcoming publication~\cite{BV}.

As shown in ref.~\cite{CC1} the larger eigenvalue of the singlet
anomalous dimension matrix, $\gamma_+(N)$, may be obtained in the
$Q_0$ scheme~\cite{CIA} as the solution of
\beq
\label{eq3}
 1 = \frac{\ab}{N - 1} \left [ \chi_0(\gamma_+)
     + \alpha_s \chi_1(\gamma_+) \right] \: ,
\eeq
where the second term is the sum of~\cite{CC1,CC2}
\begin{eqnarray}
\label{eq4}
\alpha_s \chi_1^{q\overline{q}} &\!\! =\!\! & \frac{N_f \alpha_s}{6\pi}
 \left[\frac{1}{2} \left ( \chi_0^2(\gamma)  + \chi'_0(\gamma) \right)
 - \frac{5}{3} \chi_0 - \frac{1}{2 N_c^2} \left( \frac{\pi}{\sin(\pi
 \gamma)} \right)^2 \frac{3 \cos(\pi \gamma)}{1 - 2 \gamma} \frac{2 +
 3\gamma(1 - \gamma)}{(1 + 2\gamma)(3 - 2 \gamma)} \right ] \\
\alpha_s \chi_1^{gg} &\!\! =\!\! & \frac{C_A \alpha_s}{4\pi}
 \left[ - \frac{11}{6} \left ( \chi_0^2(\gamma)  + \chi'_0(\gamma)
 \right) + \left (\frac{67}{9} - \frac{\pi^2}{3} \right) \chi_0 + \left
 ( 6 \zeta(3) + \frac{\pi^2}{3\gamma(1 - \gamma)} + \tilde{h}(\gamma)
 \right) \right.  \nonumber\\ & &~~ \left.
 \quad\quad\mbox{}-\left( \frac{\pi}{\sin(\pi \gamma)} \right)^2
 \frac{ \cos(\pi \gamma)}{3(1 - 2 \gamma)} \left ( 11 + \frac{\gamma(1
 - \gamma)} {(1 + 2\gamma)(3 - 2 \gamma)} \right) \right ] \: .
\label{eq5}
\end{eqnarray}
The function $\tilde{h}(\gamma)$ in eq.~(\ref{eq5}) is given by
\beq
\label{eq6}
 \tilde{h}(\gamma) \simeq \sum_{k=1}^3 a_k \left( \frac{1}{k+\gamma} +
 \frac{1}{1+k-\gamma} \right)
\eeq
with $a_1 =0.72, a_2 = 0.28 $ and $a_3 = 0.16$ \cite{CC2}. From these
results the irreducible NL$x$-contribution to $\gamma_{gg}$ is then 
inferred by \cite{CC3}
\beq
\label{eq7}
 \gamma_{gg}^{(1)} - \frac{\beta_0}{4\pi}\alpha_s^2 \,\frac{d}
 {d\alpha_s} \ln \left(\gamma_{gg}^{(0)} \sqrt{-\chi_0'(\gamma_{gg}
 ^{(0)})} \,\right) = \gamma^{+(1)} - \frac{C_F}{C_A} \gamma_{qg}^{(1)}
 \equiv - \frac{\alpha_s \chi_1(\gamma_{gg}^{(0)})}{\chi_0'(\gamma_{gg}
 ^{(0)})} - \frac{C_F}{C_A} \gamma_{qg}^{(1)} \: .
\end{equation}

Our subsequent numerical analysis will be performed in the DIS
factorization scheme. Here $\gamma_{gg}^{(1)}$ is represented as
\begin{eqnarray}
\label{eq8}
 \gamma_{gg,\rm DIS}^{\rm (1)} &=& \gamma_{gg,\, Q_0}^{(1)}
 + \frac{\beta_0}{4\pi} \alpha_s^2 \,\frac{d\ln R(\alpha_s)}{d\alpha_s}
 + \frac{C_F}{C_A} \left[1-R(\alpha_s)\right] \gamma_{qg,\, Q_0}^{(1)}
\\
&=& \alpha_s \sum_{k=1}^{\infty} \left[ \frac{N_f}{6\pi} \left(
 d_{gg,k}^{\, q\overline{q},(a)} + \frac{C_F}{C_A}
 d_{gg,k}^{\, q\overline{q},(b)} \right ) + \frac{C_A}{6\pi}
 d_{gg,k}^{\, gg} + \frac{\beta_0}{4\pi} \hat{r}_k \right ] \left
 (\frac{\ab}{N-1} \right)^{k-1}
\nonumber\\
&\equiv & \overline{\alpha}_s \sum_{k=0}^{\infty} b^{\, g,(1)}_k
 \left( \frac{\overline{\alpha}_s}{N-1} \right)^{k-1},
\label{eq9}
\end{eqnarray}
with $R(\alpha_s)$ defined in ref.~\cite{CH} and $\beta_0 = (11/3)\,
C_A - (2/3)\, N_f$.
Tables of the expansion coefficients $d_{gg,k}^{\, q\overline{q},(a,b)}
$, $d_{gg,k}^{\, gg}$ and $\hat{r}_k$ may be found in ref.~\cite{BV}.
Here we list for brevity only the numerical values of the first 15
coefficients $b_k^{\, g,(0)}$ and $b_k^{\, g,(1)}$ for the L$x$ and
NL$x$ series for $N_f = 4$, see Table~1. Note that the new terms
$b^{\, g,(1)}_{\, 0,1}$ agree with the corresponding results from
fixed-order perturbation theory already taking into account the
`irreducible' part of $\gamma_{gg}^{(1)}$ only.
Collecting all presently available information, the anomalous 
dimensions for the resummed unpolarized singlet evolution in the DIS
scheme are given by
\bea
\label{eq10}
\GAM(N,\ab )_{\rm DIS} &\! =\! &
 \ab \GAM_0 (N) + \ab^2 \GAM_1 (N)_{\rm DIS} \\
& & \mbox{}
 + \sum_{k=2}^{\infty} \left( \frac{\ab}{N\! -\! 1} \right)^{k+1}
   \left[
   \left( \begin{array}{cc} \! 0            & \! 0     \! \\
                            \! C_F/C_A \,   & \! 1   \!
   \end{array}\right) b_k^{\, g,(0)}
 + (N\! -\! 1)
   \left( \begin{array}{cc}
    \! C_F/C_A \: b_k^{q,(1)} & \! b_k^{\, q,(1)} \! \\
    \! 0                      & \! b_k^{\, g,(1)} \!
   \end{array} \right) \right] \:\: . \nonumber
\eea
Here $\GAM_0$ and $\GAM_1$ denote the leading and next-to-leading
order singlet anomalous dimension matrices.

\begin{table}[ht]
\begin{center}
{\small
\begin{tabular}{||r||r|r|r||}
\hline\hline
& & & \\[-4mm]
\multicolumn{1}{||c||}{$k$} &
\multicolumn{1}{c|}{$b^{\, g,(0)}_k$} &
\multicolumn{1}{c|}{$b^{\, g,(1)}_k$} &
\multicolumn{1}{c||}{$b^{\, g,(1)}_k /\, b^{\, g,(0)}_k$} \\
& & & \\[-4mm] \hline\hline
& & & \\[-4mm]
    0 &  1.000$\,$E+00 &  $-$1.139$\,$E+00   &  $-$1.14\hspace*{5mm} \\
    1 &  0.000$\,$E+00 &  $-$8.519$\,$E$-$01 &                       \\
    2 &  0.000$\,$E+00 &     3.167$\,$E$-$01 &                       \\
    3 &  2.404$\,$E+00 &  $-$1.166$\,$E+01   &  $-$4.85\hspace*{5mm} \\
    4 &  0.000$\,$E+00 &  $-$9.104$\,$E+00   &                       \\
    5 &  2.074$\,$E+00 &  $-$1.554$\,$E+01   &  $-$7.49\hspace*{5mm} \\
    6 &  1.734$\,$E+01 &  $-$1.511$\,$E+02   &  $-$8.71\hspace*{5mm} \\
    7 &  2.017$\,$E+00 &  $-$1.350$\,$E+02   & $-$66.95\hspace*{5mm} \\
    8 &  3.989$\,$E+01 &  $-$4.513$\,$E+02   & $-$11.31\hspace*{5mm} \\
    9 &  1.687$\,$E+02 &  $-$2.226$\,$E+03   & $-$13.19\hspace*{5mm} \\
   10 &  6.999$\,$E+01 &  $-$2.533$\,$E+03   & $-$36.19\hspace*{5mm} \\
   11 &  6.613$\,$E+02 &  $-$1.006$\,$E+04   & $-$15.21\hspace*{5mm} \\
   12 &  1.945$\,$E+03 &  $-$3.540$\,$E+04   & $-$18.20\hspace*{5mm} \\
   13 &  1.718$\,$E+03 &  $-$5.245$\,$E+04   & $-$30.54\hspace*{5mm} \\
   14 &  1.064$\,$E+04 &  $-$2.060$\,$E+05   & $-$19.35\hspace*{5mm} \\
[1mm]\hline \hline
\end{tabular}
\normalsize
}
\end{center}

\vspace{2mm}
\noindent
{\sf Table~1:~~The expansion coefficients $b_k^{\, g,(0)}$ and
 $b_k^{\, g,(1)}$ for the small-$x$ resummed anomalous dimension
 $\gamma_{gg}$. The latter quantities are given for four active
 flavors.  For comparison to previously employed estimates \cite{BV1}
 also the ratios of these coefficients are shown.}
\end{table}

The resummed terms beyond $O(\alpha_s^2)$ in eq.~(\ref{eq10}) do not
comply with the energy--momentum sum rule for the parton densities,
which requires
\beq
\label{eq11}
  \gamma_{qq}(N,\alpha_s ) + \gamma_{gq}(N,\alpha_s ) = 0 \: , \:\:\:
  \gamma_{qg}(N,\alpha_s ) + \gamma_{gg}(N,\alpha_s ) = 0  \: .
\eeq
This relation is satisfied by the fixed--order anomalous dimensions
order by order in $\alpha_s$. The method to restore the sum rule with
the least impact on the small-$x$ results is a (diagonal) subtraction
at $N = 2$, i.e., the addition of appropriate $\delta (1-x)$
contributions to the higher-order quark--quark and gluon--gluon 
splitting functions. We will label this prescription as ($A$) below. 
Other possibilities are the inclusion of somewhat less singular 
$1/(N-1)$ terms, later to be superseded by explicit calculations. In 
this manner a rough estimate can be obtained of the possible effect of 
subleading small-$x$ contributions to the higher-order anomalous 
dimensions, particularly in those cases where only the first term is 
known currently. We will illustrate this procedure by the prescription 
($D$) given by
\beq
\label{eq12}
  \gamma_{ij}(N) \rightarrow \gamma_{ij}(N) \, (1 - 2[N\! -\! 1]
  + [N\! -\! 1]^3) \:\:\: \mbox{ for }~ ij~=~qq, qg, gq \: .
\eeq
The terms $\propto [N\! -\! 1]$ in $\gamma_{gg}$ are taken from 
eq.~(\ref{eq9}), hence only the terms $\propto [N\! -\! 1]^3$ in this 
quantity are adjusted according to eq.~(\ref{eq11}).

We are now ready to discuss the numerical effects of the small-$x$
resummations. For definiteness, we choose the initial distributions
of the MRS(A$'$) global fit~\cite{MRSA}. Both the gluon and the sea
quark densities behave as $x^{-0.17}$ for $x \!\rightarrow\! 0$ at the
starting scale of $Q_0^2 = 4 \mbox{ GeV}^2$. The evolution is performed
for four massless flavors, and also $\Lambda^{(4)} = 231 $ MeV is
adopted from the MRS analysis. We stress that these results are mainly
theoretical illustrations. Detailed data analyses would require some
flexibility of the input gluon density at small $x$, which is only
indirectly constrained by structure function data, as well as the
inclusion of heavy-flavor mass effects.

Figure~1 displays the evolution of the proton singlet quark and gluon
(momentum) distributions, $x\Sigma(x,Q^2)$ and $xg(x,Q^2)$, in the DIS
scheme. Different resummation approximations are compared with the NLO
results (full lines). The results with the new $b_k^{\, g,(1)}$ terms 
omitted are marked by NL$x_q$.

In the quark sector the resummation corrections using prescription
($A$) are very large, e.g., they exceed a factor of four at $x=10^{-5}$
and $Q^2=100~\GeV^2$. This huge correction is entirely dominated by the 
quarkonic (upper row) anomalous dimensions. Omitting the $b_k^{\, 
g,(1)}$ contributions, and even ignoring the L$x$ terms (that case is 
not shown in the figure), has an impact of less than about 20\% on the 
singlet distribution~\cite{STU1,BV1}. The quark evolution is, however, 
very much affected by possible subleading contributions to $\gamma_{qg}
$ and $\gamma_{qq}$, even by such terms which are smaller than those 
now found for $\gamma_{gg}$. This is illustrated by prescription ($D$), 
where the less singular pieces actually overcompensate the effect of the 
leading $l=1$ contribution.

The effect of the new contribution $\gamma_{gg}^{(1)}$ to the anomalous 
dimension is, on the other hand, very substantial for the gluon 
density: the results even fall noticeably below the NLO evolution, and 
also below our previous lower estimate ($D$) at NL$x_q$ accuracy
\cite{BV1}. There is no convergence so far, but the analysis of the 
known LO and NLO results supports some hope that the inclusion of also 
two further subleading terms (series), $l=2$ and $l=3$, may lead to a 
sufficiently stable result~\cite{BV}. At this point the question arises 
how large a correction the presently unknown resummed $\gamma_{gq}
^{(1)}$ term may introduce. Experience in NLO and the L$x$ and NL$x_q$ 
resummations suggests that the impact of this contribution is of the 
order of 10\% or less, for the evolution of both $x\Sigma$ and $xg$.

We now turn to the proton structure functions. As we are working in 
the DIS scheme, $x\Sigma(x,Q^2)$ already reflects $F_2$ up to the
non-singlet pieces which are not relevant at small $x$~\footnote{The
resummation of the small-$x$ terms (L$x$) for the non-singlet structure
functions was performed in \cite{BV3}. The corresponding corrections
are smaller than 1\% over the whole $x$ range.}. Hence we  directly
turn to $F_L(x,Q^2)$. The resummed results, employing the parton
densities shown in Figure~1 and the $l=1$ resummation of the
coefficient functions \cite{CH},
\beq
\label{eq13}
  C_L(N) = \ab \, C_L^{\, 0} (N) + \ab^2 \, C_L^{\, 1} (N)_{\rm DIS}
  + \sum_{k=2}^{\infty} c_{L,k}\, \ab \left(\frac{\ab}{N-1}\right)^k
  \: ,
\eeq
are depicted in Figure~2(a). Here $C_L^{\, 0}$ and $C_L^{\, 1}$ stand
for the leading and next-to-leading order~\cite{ZN} coefficient
functions. Note that a full NL$x$ calculation of $F_L(x,Q^2)$ requires
the knowledge of the presently unknown next--order resummed coefficient
function even in the DIS scheme.

The resummation corrections are exceedingly large at the lower $Q^2$
values shown. Here they are entirely dominated by the resummed
coefficient functions. At very small values of $x$ the resummation does
even violate the condition $F_L\leq F_2$ at $Q^2\simeq 4\mbox{ GeV}^2$,
thus requiring more terms in the resummation or an adjustment of the
input densities at $Q^2_0$. At high $Q^2 \gsim 100 \mbox{GeV}^2$, due
to the decrease of $\alpha_s$ and the parton evolution, the effects of
the quarkonic anomalous dimensions and the coefficient functions are
of the same order. However, besides the anomalous dimensions also the
coefficient function will receive subleading corrections, which are
presently unknown.
To estimate the possible consequences of these terms in $C_L$ mentioned 
above, Figure~2(b) repeats the calculation illustrated by Figure~2(a), 
but with an estimate for those unknown contributions $C_L \rightarrow 
C_L (1 - 2 [N - 1])$ beyond next-to-leading order. As in the quark 
evolution, already a moderate correction can lead to an even drastic 
overcompensation of the $l=1$ effect, calling for the evaluation of the 
next resummation contributions to $C_L$.

Let us summarize: Recent results in refs.~\cite{CC1,CC2} for the first 
time allow a determination of a subleading small-$x$ resummed anomalous
dimension, namely the `irreducible' part of $\gamma_{gg}$ in the
unpolarized case. We have extracted the corresponding expansion 
coefficients in the usual DIS scheme. The coefficients of the 
subleading small-$x$ poles turn out to be mainly of opposite sign than 
the L$x$ pieces, and they are typically much larger in $N$-space.
The numerical impact of these additional terms on the proton's parton
densities and structure functions has been studied. It is largest for
the gluon evolution, where a substantial overcompensation of the
positive leading resummation effect takes place, but less important
for $x\Sigma$ and $F_L$, which are dominated by the quarkonic anomalous
dimensions and the coefficient functions. Note, however, that the 
`energy--dependent' contributions still need to be derived. All in all, 
more terms need to be calculated in the small-$x$ expansions, both of 
the anomalous dimensions and the coefficient functions, in order to 
arrive at stable resummation predictions.

\vspace{3mm}
\noindent
{\bf Acknowledgements.}
We would like to thank M. Ciafaloni, L. Lipatov, S. Catani, and 
G.~Camici for useful discussions. This work was supported in part by 
the German Federal Ministry for Research and Technology (BMBF) under 
contract No.\ 05~7WZ91P~(0).
\vspace*{-1mm}


\newpage
\centerline{\epsfig{file=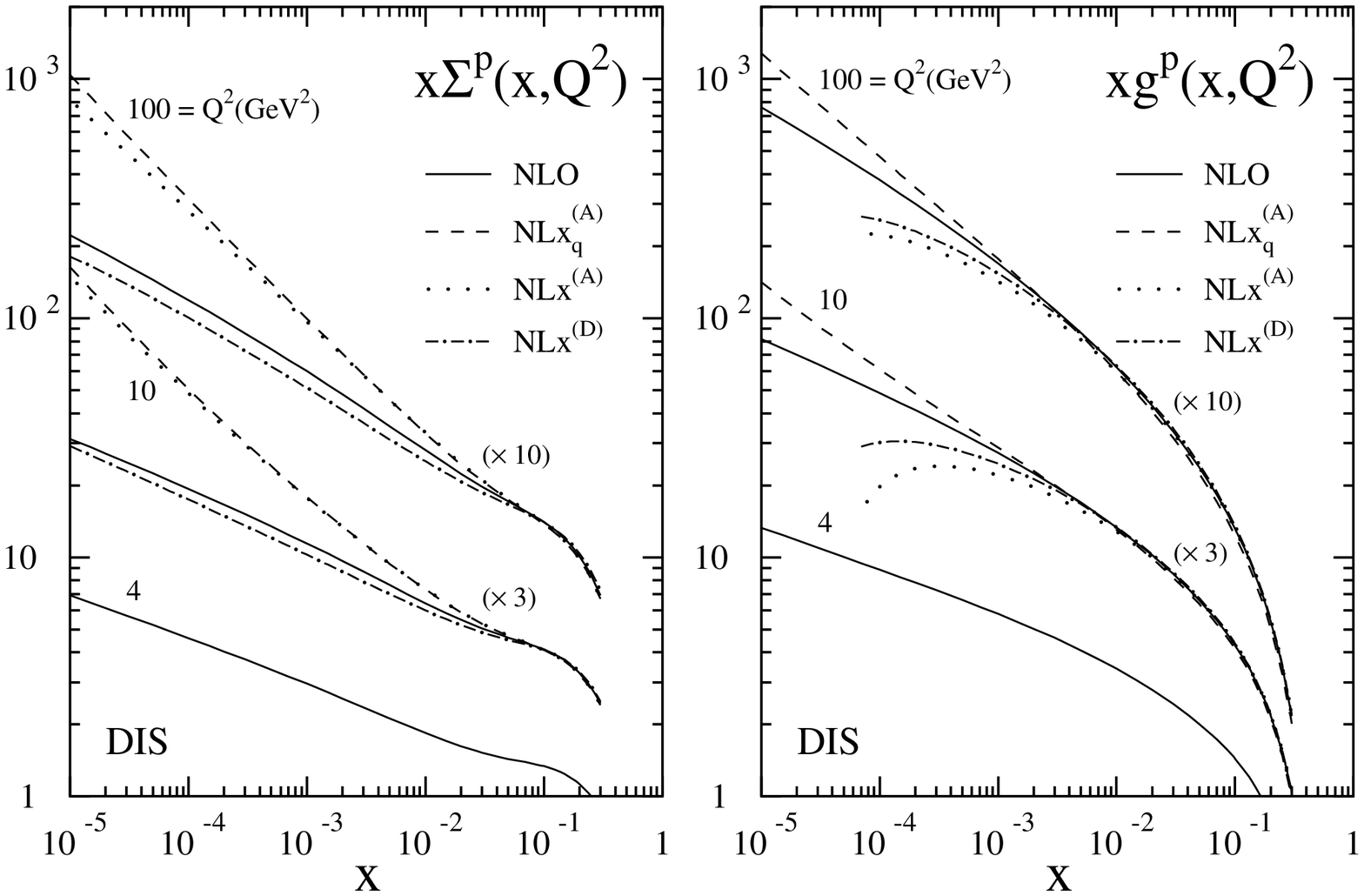,width=15cm}}
\noindent
{\sf Fig.\ 1:~~The small-$x$ evolution of the singlet quark and gluon
 densities including the resummed NL$x_q$ kernels \cite{CH} and the 
 new terms of the gluon--gluon anomalous dimension \cite{CC1,CC2} as 
 compared to the NLO results. Two prescriptions for implementing the 
 momentum sum rule have been applied.}

\centerline{\epsfig{file=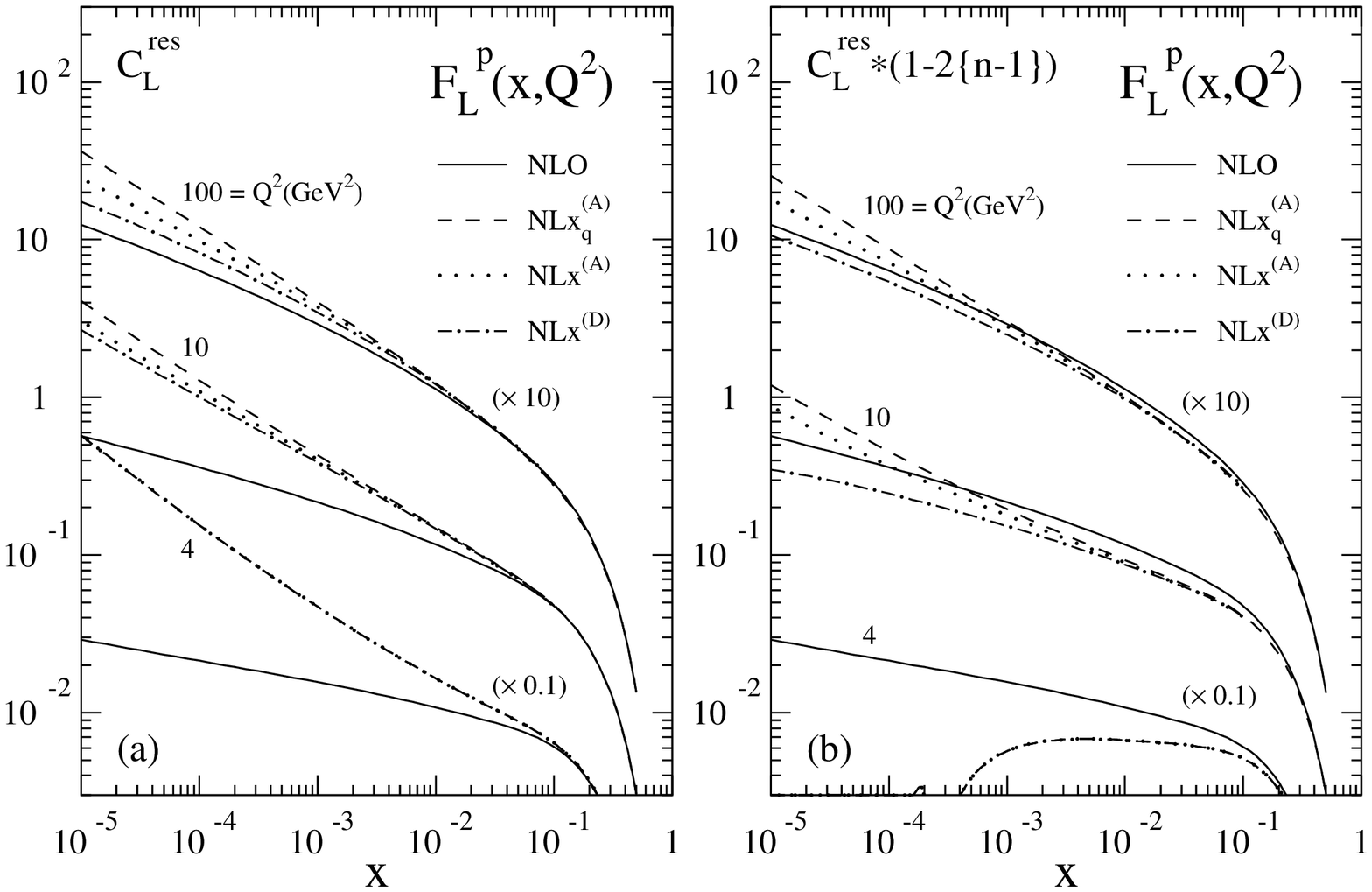,width=15cm}}
\noindent
{\sf Fig.\ 2:~~The small-$x$ behavior of the longitudinal structure
 function for the parton distributions shown in Fig.~1. The
 resummed coefficient functions of ref.\ \cite{CH} have been employed,
 in (a) without, in (b) with a moderate subleading contribution.}

\end{document}